# *A Tale of Two Metals*:

# contrasting criticalities in the pnictides and hole-doped cuprates*

N. E. Hussey, J. Buhot and S. Licciardello

*High Field Magnet Laboratory (HFML-EMFL), Institute for Molecules and Materials,*

*Radboud University, Toernooiveld 7, 6525 ED, Nijmegen, Netherlands.*

The iron-based high temperature superconductors share a number of similarities with their copper-based counterparts, such as reduced dimensionality, proximity to states of competing order, and a critical role for $3d$ electron orbitals. Their respective temperature-doping phase diagrams also contain certain commonalities that have led to claims that the metallic and superconducting properties of both families are governed by their proximity to a quantum critical point (QCP) located inside the superconducting dome. In this review, we critically examine these claims and highlight significant differences in the bulk physical properties of both systems. While there is now a large body of evidence supporting the presence of a (magnetic) QCP in the iron pnictides, the situation in the cuprates is much less apparent, at least for the end point of the pseudogap phase. We argue that the opening of the normal state pseudogap in cuprates, so often tied to a putative QCP, arises from a momentum-dependent breakdown of quasiparticle coherence that sets in at much higher doping levels but which is driven by the proximity to the Mott insulating state at half filling. Finally, we present a new scenario for the cuprates in which this loss of quasiparticle integrity and its evolution with momentum, temperature and doping plays a key role in shaping the resultant phase diagram.

*This key issues review is dedicated to the memory of Dr. John Loram whose pioneering measurements, analysis and ideas inspired much of its content.

## 1. INTRODUCTION

The discovery of the copper-oxide high temperature (high-$T_c$) superconductors in the mid-eighties [Bednorz86] was greeted with much fanfare and no small amount of hype. The hype was understandable, if premature. Suddenly, the unique properties of superconductors - from loss-free transmission of electrical power, through levitated trains to the ultimate sensitivity in brain imaging - could all be accessed using liquid nitrogen, a cheap and abundant cryogen. But while the rise in the maximum $T_c$ during the first decade of high-$T_c$ research was spectacular, the 'brute-force' approach of elemental substitution eventually ran out of steam and the promise of room temperature superconductivity – and the technological revolution that it would bring - evaporated with it. Despite this setback, the field continued to flourish, fuelled by the realization that the 'strange' metallic state out of which high-$T_c$ superconductivity emerges is unlike any other that had been seen before. Indeed, the core mysteries of the cuprates - the origin of electron pairing, the normal state pseudogap and the strange metal - are cornerstones of modern condensed matter physics that have engaged many of the world's leading theorists and acted as a catalyst for the development of experimental and theoretical tools of unprecedented sophistication. Yet the stark failure to describe strange metals within conventional theories has led to a consensus that they are fundamentally new states of matter that require an entirely new theoretical framework.

Before the discovery of superconductivity in the iron pnictides in 2006 [Kamihara06], the high-$T_c$ cuprates were considered a class apart. The former's own discovery was no less surprising – superconductivity in a compound in which the iron $d$-shell was playing a decisive role – and seemed to confirm a growing consensus that the original guiding principles for the optimisation of superconductivity, devised by Bernd Matthias in the 1960's, were now obsolete. Instead of searching for new materials with high (cubic) symmetry, no neighbouring magnetic or insulating behaviour and no oxygen, high-$T_c$ superconductivity was now the preserve of oxide materials with low dimensionality, on the verge of an insulating state and/or magnetic order [Mazin10]. The new guiding principle, demonstrated elegantly too in the heavy fermions by Lonzarich and co-workers [Mathur98], was that favourable conditions for robust superconductivity could be found close to a magnetic QCP where the magnetic order is tuned to zero by a non-thermal parameter such as chemical composition or pressure.

Standing the cuprates and pnictides shoulder-to-shoulder, there is certainly a number of striking similarities between the two families. Firstly, it is widely believed that strong electron correlations are largely responsible for both the normal-state properties and the superconductivity in these systems. Moreover, the superconductivity in both the cuprates and the pncitides arises at doping concentrations close to where antiferromagnetism (AFM) is the stable ground state, while $T_c$ itself is dome-shaped, falling away as either system is doped and becomes more itinerant. Finally, near optimum doping, many normal-state properties exhibit a marked deviation from conventional Fermi liquid behaviour. Structurally, iron pnictides also share some commonality with the cuprates, particularly their low dimensionality that is also reflected in their electronic properties. Pnictides comprise stacks of alternating Fe-pnictogen and reservoir layers, the latter donating charge to the former or in some cases, creating internal pressure.

There are important differences too, however. While the parent compounds of the iron pnictides are found to be correlated (Hund) metals, the half-filled cuprates are Mott insulators, implying that the ratio of the on-site repulsive interaction to the (non-interacting) band width is significantly larger in the latter. Moreover, whereas in cuprates, the essential physics is captured by a single Cu *d* orbital hybridised with the oxygen *p* states, the pnictides have six electrons occupying the nearly degenerate 3*d* Fe manifold, implying that in the Fe-based superconductors, inter-orbital Coulomb effects are significant, playing a more critical role perhaps than the on-site Coulomb repulsion.

As will be described below, the case for superconductivity mediated by quantum critical (spin) fluctuations in at least one member of the pnictide family is now overwhelming, if not yet universally confirmed. In the cuprates, however, the situation is still shrouded in controversy. Superconductivity in the high-$T_c$ cuprates is most robust (i.e. its superfluid density is maximised) at a doping level that coincides with the closure of the so-called pseudogap, a momentum-selective depletion of spectral weight which, despite sporadic, unsubstantiated reports in other materials, appears to be unique to the hole-doped cuprates. In quantum critical models for cuprate superconductivity, the pseudogap is associated with some form of symmetry-breaking. Indeed, over the past few years, there have been numerous experimental observations of electronic ordering phenomena in lightly doped cuprates that have raised many interesting questions regarding the role of fluctuating order both in mediating pairing and in the formation of the pseudogap. Intriguingly, however, the evolution of the transport and thermodynamic properties, often the most revealing probes of quantum criticality, do not support such a conventional quantum critical scenario.

This review will not feature a long, detailed exposé on all the physical manifestations of the pseudogap. While this is a crucial element in the contrasting pictures that emerge for the two families, it is not the main focus here and in any case, there are already a number of excellent reviews on the topic in the current literature [Timusk99, Tallon01, Hufner08, Kordyuk15]. Rather, we will focus on the essential differences in the bulk transport and thermodynamic properties of the two families, ultimately reaching the conclusion that while there are indeed critical points associated with charge or spin ordering in underdoped cuprates, they account for neither the pseudogap nor the strange metal physics observed on the overdoped side. We conclude the review with speculation as to what then is driving pseudogap formation in cuprates and propose an alternative scenario for the cuprates, one in which the loss of quasiparticle integrity plays the defining role.

We start with a short introduction to quantum criticality in Section 2, and in Section 3, we introduce the concept of quasiparticle integrity in correlated metals and the bounds beyond which this integrity is lost. Section 4 then summarises the evidence for quantum critical behaviour in the iron pnictide family $BaFe_2(As_{1-x}P_x)_2$ which will be used as a reference point when discussing the evidence for electronic ordering phenomena in hole-doped cuprates in Section 5. In Section 6, we describe the relevant transport and thermodynamic properties of hole-doped cuprates that contrast markedly with what is found in $BaFe_2(As_{1-x}P_x)_2$ and which motivate us to introduce an alternative scenario for the cuprate phase diagram, based on the loss of quasiparticle integrity at different loci on the original, coherent Fermi surface. Finally, in Section 7, we discuss some of the implications of this new approach and conclude.

## 2. QUANTUM CRITICALITY

The phenomenon of quantum criticality is associated predominantly, though not exclusively, with the breaking of some form of symmetry (e.g. translational, rotational or time-reversal symmetry) within a crystalline lattice. While quantum phase transitions are most commonly found in magnetic insulators, we focus here, for obvious reasons, on correlated metals that can be tuned through a QCP. For strongly correlated electrons, a QCP is obtained when either (i) the long-range order is suppressed to $T = 0$ (second-order phase transition) or (ii) the critical end-point terminating a line of first-order transitions is depressed to $T = 0$. At finite temperatures, a phase transition occurs as a spontaneous symmetry breaking, flanked by critical fluctuations of the relevant order parameter (e.g. spin, charge or orbital order). A non-thermal tuning parameter such as pressure, chemical doping, or geometrical frustration, can then be used to enhance the itinerancy of key electronic states which in turn suppresses the transition temperature. When the ordering temperature is sufficiently low, long-wavelength critical fluctuations of the order parameter become quantum mechanical in nature and through their coupling to the low-energy quasiparticle excitations around the Fermi surface, give rise to novel non-Fermi liquid physics.

Figure 1a shows a typical phase diagram of a quantum critical system. As temperature is lowered towards a continuous (second-order) phase transition at $T_{c1}$, fluctuating droplets of the impending order begin to form and swell as the system is tuned to the critical point $g_c$. These spatial fluctuations are characterized by a diverging correlation length $\xi \propto |g - g_c|^{-\nu}$ whose rate of divergence is set by the anomalous exponent $\nu$. At the critical point, droplets of all sizes permeate the entire material.

In a quantum phase transition, dynamical fluctuations of the order parameter also play a role, in addition to the usual spatial fluctuations, and are allowed even at zero temperature [Hertz76]. These dynamical (temporal) fluctuations are characterized by a correlation time $\tau$ that too diverges, but as $|g - g_c|^{-z\nu}$, where, $z$ is the dynamic exponent of the quantum phase transition. As $g$ increases, $T_{c1}$ decreases. When the (quantum) fluctuations are strong enough, the system undergoes a transition to a new ground state without any change in entropy. The locus of this zero-temperature axis of the phase diagram defines the QCP. At this point, the states on either side of $g_c$ are intertwined, and the wave function becomes an entangled quantum superposition of the two.

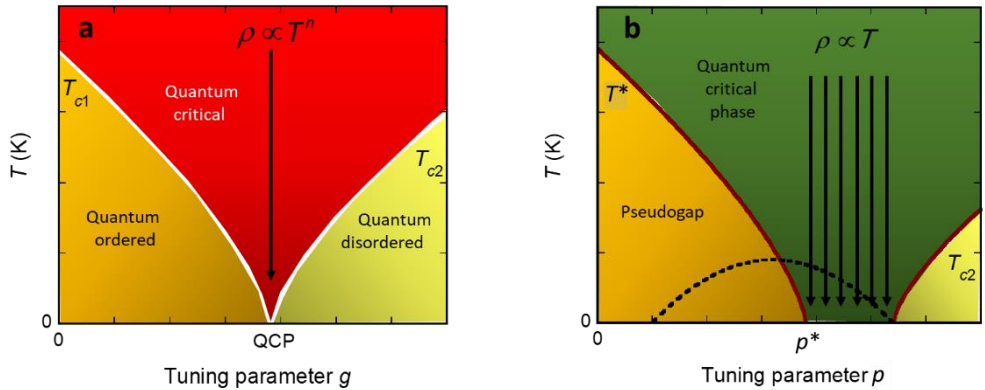

**Figure 1**: Comparison of the phase diagram for a) a conventional quantum critical metal and b) a putative strange metal. In contrast to a), where a phase transition at $T = T_{c1}$ is tuned to zero temperature at a quantum critical point (QCP) separating the quantum ordered from a quantum disordered phase, in b) an anomalous quantum critical *phase* exists over a region of the phase diagram possessing different scaling dimensions that can change continuously.

In addition to the correlation time, there is also a thermal timescale $L_\tau = \hbar/k_B T$ that characterizes the thermal fluctuations. In the low-temperature regime where $L_\tau \gg \tau$, electronic excitations are only weakly influenced by thermal fluctuations, the quasiparticles are well defined and the physical properties are Fermi-liquid-like. Approaching the QCP along the horizontal axis, however, the quasiparticle scattering cross-section grows as the fluctuations of the associated order parameter slow down and occur over increasingly long wavelengths. This in turn leads to a divergence in the coefficients of the $T^2$ resistivity and the electronic specific heat. In the high-$T$ regime, however, $L_\tau \ll \tau$ and striking deviations from Fermi-liquid behavior can be observed, such as $T$-linear resistivity (over an extended temperature range) and a logarithmically divergent specific heat coefficient $\gamma(T)$. The overall dependence of $L_\tau$ and $\tau$ give the quantum critical region a characteristic fan shape, the importance of quantum criticality expanding with increasing $T$ far beyond the isolated QCP at $T = 0$.

The main motivation for studying itinerant systems near a putative QCP is that it provides an opportunity to investigate thermodynamic and transport properties exhibiting behaviour that departs significantly from conventional Fermi-liquid theory. The precise form of each physical property depends on a number of factors, including the type of order associated with the QCP and the dimensionality of the electronic state. Nonetheless, the phase diagrams of candidate quantum critical systems are, on the whole, rather generic.

## 3. LOSS OF QUASIPARTICLE INTEGRITY

As discussed above, well-defined quasiparticles are believed to exist below the fan of quantum criticality depicted in Figure 1a. Being a central theme of this article, it is important that we formulate carefully our definition of a well- or ill-defined quasiparticle. While quasiparticles are only approximate eigenstates of an interacting many-body system, their properties resemble very closely those of their non-interacting counterparts, albeit with renormalized physical parameters such as the modified effective mass $m^*$ (relative to its band mass $m_b$). Related to $m^*$ is the quasiparticle weight $Z$ – the (normalized) spectral weight of the quasiparticle peak at the Fermi energy $E_F$. This peak encompasses the coherent excitations that form the Fermi liquid and are responsible for the Drude optical response. Its spectral weight decreases as the (electron-electron) interaction strength is turned up and/or as the metal-insulator transition is approached. Under certain simplifying assumptions, $m^*/m_b = 1/Z$.

The fate of a quasiparticle is closely linked with its decay rate, i.e. the rate at which it loses phase memory. Quasiparticles must retain their local quantum phase coherence over a timescale $\tau_\varphi$ that is longer than the thermal scale $L_\tau$ ($\tau_\varphi$ should not be confused with $\tau$, the correlation time for dynamic fluctuations of an order parameter). According to Fermi-liquid theory, a quasiparticle is always well-defined (i.e. it is a long-lived eigenstate) near the Fermi surface as its decay rate (~ $E^2$) always becomes smaller than its excitation energy *E*. As the temperature is increased, however, the adiabatic continuity linking the quantum numbers (labels) of the quasiparticle with those of its non-interacting counterpart breaks down and the quasiparticle will decay before the interaction is completely turned on, implying that its integrity is lost. Thus, above a characteristic temperature $T_{coh}$, one sees first a departure from canonical Fermi-liquid behaviour and eventually, the destruction of the quasiparticle.

The boundaries separating these different quasiparticle states are hard to pinpoint, though there are certain guiding principles, particularly relating to the strength of the decay rate, that inform whether or not the quasiparticle description remains valid. As we have seen, for a Fermi liquid, the low-energy excitations are always coherent since their effective collision time diverges as $T \rightarrow 0$ as $1/T^2$. For a system in which the decay rate $\hbar/\tau_\varphi$ vanishes linearly with temperature, however, there is a second upper bound $\hbar/\tau \sim 2\pi k_B T$ [Maldacena15, Hartnoll16] beyond which quasiparticle coherence is lost. This boundary is often referred to as the Planckian dissipation limit [Zaanen04] and represents a crossover from transport dominated by momentum relaxation to one driven by diffusion of charge and energy.

The second boundary is the Mott-Ioffe-Regel (MIR) limit, which states that the quasiparticle mean-free path can never be less than the lattice spacing *a*, or alternatively, its scattering rate $\hbar/\tau_\varphi$ can never exceed the effective bandwidth *W*. Metals that violate this limit are known as bad metals and are characterized by an incoherent electron spectral function, a collapse of the zero-frequency Drude peak in the optical conductivity and a thermally-driven shift of the associated spectral weight to higher frequencies [Rozenberg95, Merino00, Limelette03, Hussey04]. It is this shift of low-frequency spectral weight with increasing temperature that gives the appearance of metallic (dc) transport, rather than an ever escalating scattering rate. Indeed, the notion of a scattering rate, and by association, a quasiparticle, is no longer meaningful beyond the MIR limit.

So, in essence, there are two limits for the quasiparticle description and for quasiparticle-dominated transport: one associated with the magnitude of the scattering rate (or maximal resistivity within a conventional Boltzmann transport framework) and one associated with its temperature derivative [Hartnoll16]. It will be argued below that both these bounds play a significant role in shaping much of the cuprate phase diagram, including the development of the normal state pseudogap, without the need to invoke an underlying order parameter.

Before discussing this in more detail, however, it is worthwhile contrasting the situation in the cuprates with what is seen in an emerging example of an archetypal quantum critical system.

## 4. QUANTUM CRITICALITY IN THE PNICTIDES: A CASE STUDY

Although there are several different iron pnictide families, their electronic and crystalline structures, as well as their phase diagrams, are all quite similar. In this report, we focus on the BaFe$_2$(As$_{1-x}$P$_x$)$_2$ family, for the express reason that this is the pnictide system for which the most convincing evidence for a QCP currently exists. At the end of this section, we also consider briefly the experimental situation in the chalcogenides. Much of the evidence for quantum criticality in BaFe$_2$(As$_{1-x}$P$_x$)$_2$ has been summarised in an excellent review by Shibauchi, Carrington and Matsuda [Shibauchi14]. In this section therefore, we will simply highlight the main findings and present some more recent material not covered in Ref. [Shibauchi14] that is relevant for our subsequent discussions.

The parent compound BaFe$_2$As$_2$ transitions at $T_N \sim$ 140K from a tetragonal paramagnet to an orthorhombic metallic antiferromagnet [Huang08], with the structural transition (broken $C_4$ symmetry) slightly preceding the magnetic one. In electron-doped Ba(Fe$_{1-x}$Co$_x$)$_2$As$_2$, the second-order nature of the transition was confirmed via the observation of a sharp peak in the electronic specific heat [Chu09]. With P substitution, the spin density wave order is rapidly suppressed and gives way to superconductivity with a maximum $T_c$ of 33 K at $x \sim$ 0.30, as shown in Figure 2a. Electronically, BaFe$_2$As$_2$ is a compensated semi-metal possessing three cylindrical hole sheets around the zone center and two electron pockets at the zone corners. Substitution of P for As leads to the loss of one of the hole sheets and a strong increase in the $c$-axis warping of another, thereby reducing the overall nesting tendencies between the different Fermi surfaces. It is claimed that P substitution also reduces the relative strength of the electron-electron correlations, the enhanced electron itinerancy providing a second pathway to reduce the effective magnetic coupling and thus drive the system across the magnetic QCP [Dai09].

The first hints of quantum criticality in BaFe$_2$(As$_{1-x}$P$_x$)$_2$ were revealed in initial transport studies [Jiang09, Kasahara10], more specifically in the evolution of the exponent $\alpha$ in the relation $\rho = \rho_0 + AT^\alpha$ as a function of $x$ (represented as a colour plot of Figure 2a). At high doping ($x \geq 0.6$), the in-plane resistivity $\rho$ follows the characteristic Fermi-liquid form $\rho = \rho_0 + AT^2$. Near optimal doping ($x = 0.33$) however, $\rho$ exhibits a linear temperature dependence, together with a strongly *T*-dependent Hall coefficient and non-Kohler's scaling in the magnetoresistance [Kasahara10]. All these results were initially taken as evidence for non-Fermi liquid behavior arising from the proximity of optimally doped compounds to a QCP.

The normal state spin dynamics of BaFe$_2$(As$_{1-x}$P$_x$)$_2$ were later investigated by nuclear magnetic resonance (NMR) [Nakai10]. At high P doping, the spin-lattice relaxation rate $1/T_1T$ is nearly temperature independent, indicating the expected Korringa relation for a Fermi liquid. Again, as *x* approaches optimally doping, $1/T_1T$ develops a strong temperature dependence, indicative of increased coupling to the spin fluctuations. The Curie-Weiss temperature $\theta$ was also found to vary strongly with *x*, crossing zero at a critical concentration *x* ~ 0.3 where $T_c$ maximizes (pink dashed line in Fig. 2a). The overall behavior is consistent with that expected near a second-order AFM critical point, where the dynamical susceptibility diverges at *T* = 0 K, and indicates that AFM fluctuations continue to grow down to $T_c$ at optimal doping. A recent NMR study [Dioguardi16] has also found evidence for glassiness in the spin dynamics of BaFe$_2$(As$_{1-x}$P$_x$)$_2$, suggesting some role for disorder in shaping the critical fluctuations. Moreover, (quadrupolar) fluctuations of nematic order have also been reported. We return to discuss the possible role of nematicity at the end of this section.

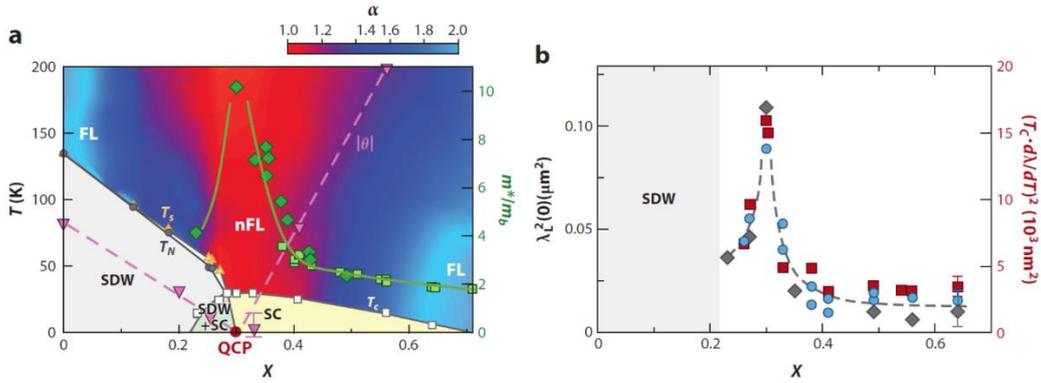

**Figure 2**: (a) Phase diagram of BaFe$_2$(As$_{1-x}$P$_x$)$_2$. The colour plot represents the evolution of the exponent $\alpha$ in $\rho_{ab}(T) = \rho_0 + AT^\alpha$ while the locations of the structural ($T_s$), magnetic ($T_N$) and superconducting ($T_c$) transitions are determined resistively. The Curie-Weiss temperature $\theta$ determined by NMR is also indicated (inverted triangles). The green symbols represent $m^*/m_b$ (right axis) determined by specific heat and quantum oscillation measurements. The putative QCP occurs at $x_c$ = 0.3 [Shibauchi14]. (b) The *x* dependence of the square of zero-temperature London penetration depth $\lambda_L^2(0)$ determined by various techniques [Hashimoto12].

The strong quantum fluctuations near the QCP lead to a significant modification of the quasiparticle masses. The *x*-dependence of $m^*/m_b$ for one of the electron sheets, as determined by quantum oscillation experiments, is plotted in Fig. 2a. The enhancement itself is striking. Near the SDW boundary, $m^*/m_b$ is up to five times that of the end member BaFe$_2$P$_2$. Moreover, the 'global' quasiparticle mass enhancement determined by the jump in the specific heat ($\Delta C$) at $T_c$ is found to be consistent with that obtained from quantum oscillations, suggesting that the mass enhancement is relatively uniform over the entire Fermi surface. This strong mass enhancement is also revealed in the magnetic penetration depth $\lambda_L(0)$ and in high-field transport measurements. In the clean limit, $\lambda_L^{-2}(0) = \mu_0 e^2 (n/m^*)$, where $n$ is the carrier density. Figure 2b shows the P-concentration dependence of $\lambda_L^2(0)$ in the zero-temperature limit, as determined by three different methods [Hashimoto12]. The sharp peak in $\lambda_L^2(0)$ at $x_c$ indicates that the superfluid density is minimal at this critical $x$ value and its form is consistent with the mass enhancement obtained by other techniques.

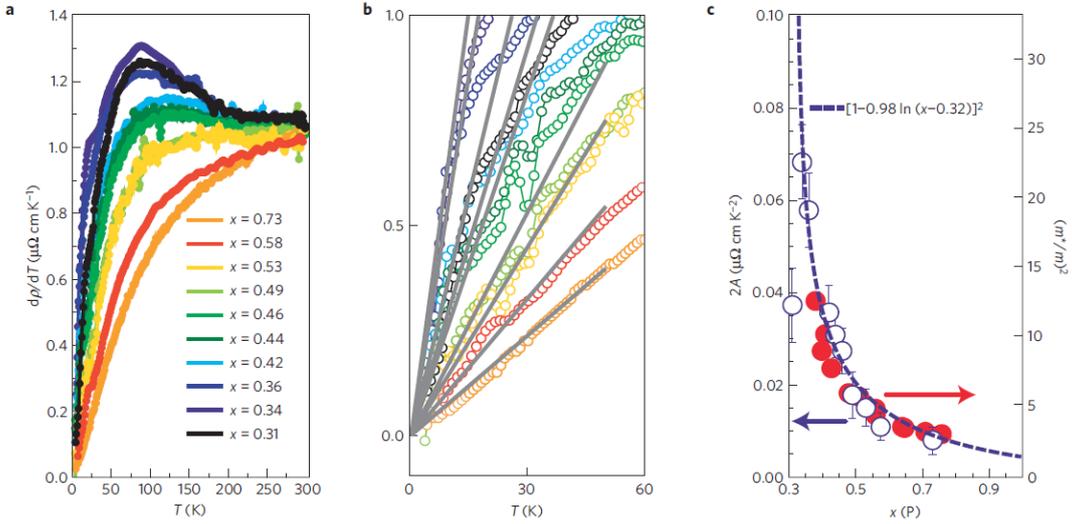

**Figure 3:** a) Derivative of the resistivity in BaFe$_2$(As$_{1-x}$P$_x$)$_2$ interpolated over the full temperature and doping range. b) Blow-up of the low-$T$ region with linear fits (grey lines). c) Variation of 2$A$ (open circles) for $0.37 \leq x \leq 0.73$ and its comparison with the square of the effective mass $m^*$ [Analytis14] The line is a fit to a phenomenological divergence of $m^*$ near a QCP $[C_0 - C_1 \ln(x-x_c)]^2$, with $x_c = 0.32$ [Abrahams11].

Figures 3a and 3b show the temperature derivative d$\rho$/d$T$ obtained on a series of longitudinal resistivity measurements on BaFe$_2$(As$_{1-x}$P$_x$)$_2$ single crystals after the application of a magnetic field strong enough to suppress the superconductivity [Analytis14]. For all $x$ values studied, the d$\rho$/d$T$ curves are linear at low temperature with a zero intercept, indicating that the limiting low-$T$ resistivity always recovers a quadratic $T$-dependence, consistent with a Fermi-liquid ground state. As optimal doping is approached, the range in $T$ over which $T^2$ behavior is observed steadily decreases. In addition, the slope of these curves (= 2$A$) diverges, as shown in Fig. 3b. Significantly, the variation of $A(x)$ is found to be in excellent agreement with that obtained by plotting the square of the quasiparticle masses obtained by quantum oscillation experiments (and by inference from Fig. 2a, also with the normal state electronic specific heat coefficient), a relation enshrined in the Kadowaki-Woods ratio first reported for heavy fermion systems [Kadowaki86, Hussey05].

The dashed blue line in Fig. 3c is a fit to a phenomenological critical power law $A \propto [C_0 - C_1 \ln(x-x_c)]^2$ expected for a quasi-2D system near an antiferromagnetic QCP [Abrahams11]. This study confirms that $BaFe_2(As_{1-x}P_x)_2$ exhibits 'conventional' quantum critical behaviour in the sense that $\rho(T)$ always crosses over from a $T$-linear to a $T^2$ dependence below a temperature scale that diminishes on approach to the QCP. In accordance with this picture, the coefficient of the $T^2$ resistivity is also found to diverge as the QCP is approached. Although there is at present no accepted theory explaining the origin of $T$-linear resistivity in quantum critical metals, a common point of view is that it is associated with Fermi-surface 'hot spots' — regions of the Fermi surface with strongly enhanced scattering. Thus, as the QCP is approached from the highly P-doped side, the Fermi-surface hotspots become increasingly 'hotter' and the non-Fermi liquid behavior indicated by the red region in Fig. 2a is most likely associated with the finite-temperature quantum critical region linked to the QCP.

In summary, $BaFe_2(As_{1-x}P_x)_2$ bears all the hallmarks of a system that can be tuned through a QCP lying beneath the superconducting dome. The transport properties, NMR, quantum oscillations, and specific heat all suggest the presence of a magnetic QCP at $x_c = 0.30$ that is responsible for the non-Fermi liquid behaviour found above $T_c$. Moreover, there is strong evidence that superconductivity and antiferromagnetism coexist in this system. What is particularly striking about these collective results is the overall consistency in the mass enhancement, despite the fact that the $m^*$ values were obtained either at zero field and low temperatures, deep in the superconducting state (via the penetration depth), at high magnetic fields and low temperatures (via quantum oscillations and high-field magnetotransport), or at finite temperatures and zero magnetic field (via the jump in the specific heat at $T_c$). This consistency in the variation of the quasiparticle masses inferred from the various probes implies that, contrary to some expectations, field-induced destruction of the superconductivity uncovers the same ground state that would be there in zero-field (in the absence of superconductivity). In addition, the fact that the highest $T_c$ is attained right at the QCP with the most enhanced mass strongly implies that the quantum critical fluctuations actually enhance superconductivity in BaFe$_2$(As$_{1-x}$P$_x$)$_2$.

As mentioned above, a recent NMR study [Dioguardi16] reported spin dynamics in BaFe$_2$(As$_{1-x}$P$_x$)$_2$ with a doping and temperature response that is consistent with the nematic susceptibility deduced, for example, from elastoresistance measurements on other Ba122 families [Chu12]. This finding suggests the presence of nematic fluctuations (which are peaked at $q = 0$), in addition to the usual spin fluctuations (peaked at $q = Q$). It has been argued that such nematic fluctuations could also play a sub-dominant but nonetheless significant role in enhancing both $m^*$ and $T_c$ in the iron pnictides near the QCP [Fernandes14, Lederer15]. The effect is believed to be stronger in two dimensions than in three [Lederer15], giving a possible explanation for the significantly enhanced $T_c$ in atomically thin films of the iron chalcogenide FeSe.

The iron chalcogenides are an important material system in this regard due to the fact that nematic order appears to exist without an accompanying magnetic ground state. Elastoresistance measurements on the isovalently substituted Fe(Se$_{1-x}$Se$_x$), for example, show a similar divergence near the ordering transition to the one found in the Ba122 family [Hosoi16], yet there is no clear evidence for magnetic correlations in this system. Thus, the nematic order in this case seems to be driven purely by orbital effects [Watson15]. It is worth noting however that the $T_c$ enhancement across the nematic QCP in Fe(Se$_{1-x}$Se$_x$) is found to be rather modest [Hosoi16]. Finally, we note that evidence for nematicity has also been reported in the cuprates [Daou09, Lawler10], and so some $T_c$ enhancement via nematic fluctuations cannot be discounted there either.

## 5. ORDERING PHENOMENA IN UNDERDOPED CUPRATES

As with the pnictides, the structural element common to all cuprates is the stack of two-dimensional conducting planes separated by charge reservoir layers. A combination of covalency, crystal field splitting and Jahn–Teller distortion then creates an electronic structure whose highest partially filled band has predominantly Cu 3$d_{x2-y2}$ and O 2$p_{x,y}$ character. At half-filling (zero hole doping), the 'parent' compounds are AFM Mott insulators with a charge transfer gap in the region of 1-2 eV. Long-range antiferromagnetism (labelled AFM in Figure 4) is destroyed with only a few percent of doped holes, though short-range AFM correlations are found to persist right to the edge of the superconducting dome (labelled SC) at high doping [Wakimoto07, LeTacon13, Dean13]. Significantly, there is no obvious change in the structure or strength of these correlations across $p^*$ (= 0.19 ± 0.01), the critical doping level that signifies the end of the pseudogap phase [Tallon01]. (Here we focus only on hole-doped cuprates for which the pseudogap is well established.) The fact that $p^*$ is located inside rather than at the edge of the SC dome argues against preformed pairs or SC phase fluctuations (whose onset is denoted $T_\varphi$ in Fig. 4) as the origin of the pseudogap. Moreover, while $p^*$ is well-defined, the corresponding $T^*$ crossover line that signifies the onset of pseudogap-related physics in different experimental techniques varies wildly and depends strongly on the criterion used.

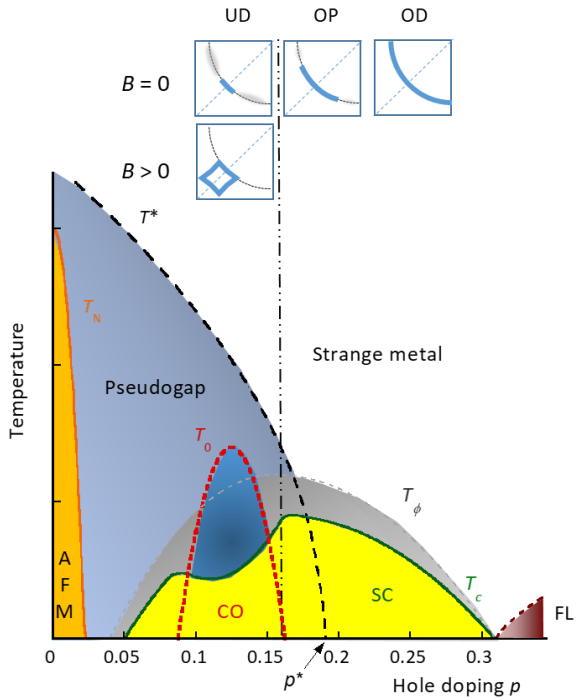

**Figure 4**: Phase diagram for hole-doped cuprates in which the pseudogap onset temperature $T^*$ collapses to zero inside the superconducting (SC) dome at a critical doping level $p^*$ ~ 0.19. Below $T_0$, short-range charge order (CO) appears as a competing phase, suppressing $T_c$. $T_\phi$ defines the extent of superconducting phase fluctuations. The set of boxes show the evolution of a quadrant of the Fermi surface (thick blue line) from the overdoped (OD) metallic side on the right, through optimal doping (OP) to the underdoped (UD) side close to the antiferromagnetic insulator (AFM). The Fermi surface first collapses into Fermi arcs then, in the presence of a magnetic field, reconstructs into electron pockets located near the zone centre.

Prior to 2007, the prevailing view of the electronic structure of underdoped and optimally-doped cuprates was one of Fermi arcs: disconnected regions of coherent quasiparticles located near the nodal region $(\pi, \pi)$ of the Brillouin zone (see Fig. 4). These arcs evolve out of an incoherent high-temperature phase and are arguably the most dramatic manifestation of the pseudogap itself. In 2007, however, a new component to the phase diagram was revealed, initially via magnetotransport and quantum oscillation experiments that indicated some form of Fermi surface reconstruction [Doiron-Leyraud07, LeBoeuf07], then later via NMR [Wu11], resonant X-ray scattering [Ghiringelli12] and sound velocity [LeBoeuf13] measurements that identified a thermodynamic phase below $T_0$ (see Fig. 4) with static, incommensurate but short-range charge order (CO), order that seemingly competes with superconductivity [Chang12].

The discovery of slow quantum oscillations in the Hall resistivity of underdoped $YBa_2Cu_3O_{6+\delta}$ (Y123) [Doiron-Leyraud07, Sebastian08] came as a major surprise, not only because it conflicted with the notion of disconnected Fermi arcs, but also because it was inconsistent with the large Fermi surface inferred from high-field magnetotransport in overdoped cuprates [Hussey03, Vignolle11]. The later observation of a sign change in both the Hall [LeBoeuf07] and Seebeck coefficients [Chang10] meant that these small pockets have electron character and that the Fermi surface is reconstructed at low temperature due to some form of translational symmetry breaking. Many proposals for the reconstruction ensued; the scenario most consistent with existing experimental data being one involving biaxial charge order linked to the tips of the remnant Fermi arcs [Harrison11, Allais14]. Subsequent detection of quantum oscillations in $YBa_2Cu_4O_8$ (Y124) [Bangura08, Yelland08] and $HgBa_2CuO_{4+\delta}$ (Hg1201) [Barisic13] along with the observation of similar transport anomalies in Hg1201 [Doiron-Leyraud13] confirmed that this reconstruction was a generic property of underdoped cuprates.

The first report of broken symmetry in the normal state of underdoped cuprates actually came with the discovery of stripes (regular patterns of antiferromagnetic domains, separated by domain walls into which the charge carriers segregate) in Nd-doped $La_{2-x}Sr_xCuO_4$ (LSCO) with $x$ = 0.12 [Tranquada95]. Subsequent transport studies of Nd-doped, Eu-LSCO and Y123 [Cyr-Choiniere09, Chang10] have also suggested that Fermi surface reconstruction in these systems is indeed caused by some form of stripe order. In underdoped Y123, on the other hand, NMR experiments revealed a magnetic-field-induced modulation of the charge density, without any sign of static magnetism [Wu11, Wu13] implying that only charge order is present in this particular system. Subsequent X-ray scattering experiments [Ghiringhelli12, Achkar12, Chang12, Blackburn13, LeTacon14] uncovered quasi-static incommensurate charge density wave (CDW) order in Y123 with domain sizes up to ~20 unit cells within the pseudogap state. Finally, the temperature and field dependence of the X-ray intensity implies a competition between (three-dimensional) CDW formation and superconductivity [Chang12, Gerber15].

The development of high-resolution scanning tunneling spectroscopy shortly after the millennium uncovered evidence of a different form of spatial modulation of electronic states, known as checkerboard order, firstly in $Bi_2Sr_2CaCu_2O_{8+\delta}$ (Bi-2212) [Hoffman02, Howald03, Vershinin04] and later in $Ca_{2-x}Na_xCuO_2Cl_2$ (Na-CCOC) [Hanaguri07]. Initially thought to signal the same stripe order that was found in the La-based cuprates, subsequent doping studies revealed that the (incommensurate) charge modulation wave vectors were linked to a charge-density-wave (CDW) order resulting from Fermi surface nesting [Wise08]. More recently, the ordering wave vector in both the Bi- and Hg-based cuprates was found to coincide with a $Q$-vector that connects the tips of the Fermi arcs, providing further support for a nesting scenario [SilvaNeto14, Comin14, Tabis14] (See Fig. 5).

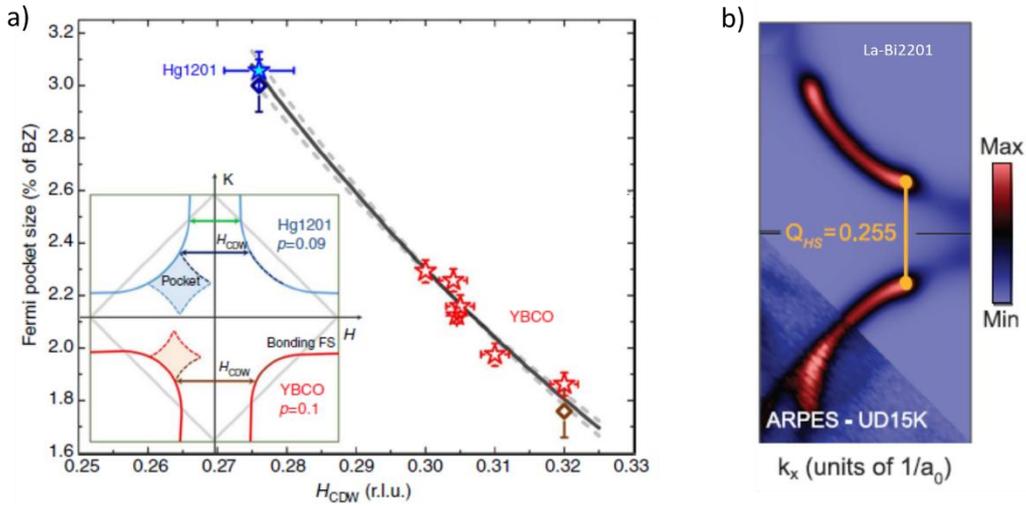

**Figure 5:** Connection between the Fermi surface topology and the ordering wave vectors (nesting) for Y123 and Hg1201 (left panel) [Tabis14] and for Bi2212 (right panel) [Comin14].

It is now widely assumed that the quantum oscillations derive from a diamond-shaped pocket located near the nodes (see inset to Fig. 5a) that is composed of the four remnant Fermi arcs translated via the incommensurate ordering wave vector. Moreover, in Bi2201, the charge ordering onset temperature was found to be comparable to $T^*$ [Comin14], suggesting that the pseudogap is related to fluctuating CDW order, though any single particle gap originating from this CDW order has not been clearly elucidated by angle-resolved photoemission (ARPES).

In the following, we will attempt to clarify some important issues regarding the relation between the pseudogap and the CDW; (i) are the quantum fluctuations associated with this charge ordering involved in the electron pairing and thus act to promote or enhance the superconductivity? (ii) Does the pseudogap define a precursor state, signaling the onset of dynamic charge modulations, or is it something distinct, a novel, correlated electronic state out of which a charge instability is nucleated? (iii) If the latter, are there other forms of order which are responsible for the pseudogap? With regards (i), quantum oscillation experiments in near optimally doped Y123 appeared to support the notion that the interactions responsible for the mass enhancement also give rise to the high-$T_c$ superconductivity [Ramshaw15]. A more recent QO study of underdoped Y124 under pressure [Putzke16], however, showed that while pressure enhances $T_c$ (and doping), the mass enhancement is in fact suppressed by pressure. The inference from this observation, then, is that quantum fluctuations of the charge order enhance $m^*$ but do *not* enhance $T_c$.

With regards (ii), Badoux *et al.* recently extended an earlier study of the Hall effect in UD Y123 ($p < 0.15$) [LeBoeuf07] by looking at samples with doping levels up to and beyond $p = 0.2$ with correspondingly higher critical fields [Badoux16]. The experiments revealed that the low-temperature sign change in the Hall effect – indicative of long-range charge order – is confined to a narrow doping range (marked as CO in Fig. 4) that terminates at or near optimal doping ($p = 0.16$). Even more significantly, they discovered a second marked change in the Hall response, namely a six-fold decrease in its magnitude, at a doping level $p \sim 0.2$ which previous thermodynamic and transport measurements had identified with the extinction of the pseudogap [Tallon01]. Indeed, the authors argue that this marked decrease (increase) in the Hall resistance (carrier density) is consistent, qualitatively if not quantitatively, with the closing of the pseudogap and the restoration of the full Fermi surface. Thus, pseudogap formation and the charge ordering terminate at different doping concentrations, implying that they stem from different origins.

Finally, if the charge ordered phase resides only inside the dome, the question then arises; what other forms of order, if any, are responsible for the pseudogap? There is currently no consensus on this important question, though equally, no shortage of contenders. In the following, we restrict ourselves to those candidates whose transition temperature is comparable with $T^*$ (as determined by bulk transport and NMR Knight shift measurements,) that decreases monotonically with increasing $p$, and most importantly, that vanishes at $p^* \sim 0.19$. $T^*$ is typically defined as the temperature below which the in-plane resistivity deviates from its high-$T$ $T$-linear dependence. While the determination of $T^*$ is sensitive to the criterion used, it appears to coincide reasonably well with the observed drop in the Knight shift $K_s$, and provided $K_s$ is assumed to be proportional to the density of states at the Fermi level, then $T*$ marks the onset of the manifestation of the pseudogap at the Fermi level $E_F$.

Figure 6 shows a sub-set of $T^*$ values as determined by polarized neutron diffraction, polar Kerr effect and resonant ultrasound measurements [Shekhter13]. Neutron diffraction studies first identified a new feature in both underdoped Y123 [Fauque06] and Hg1201 [Li08] that developed at a temperature scale found to be consistent with the emergence of second-harmonic optical anisotropy [Zhao16] and roughly 100 K above that at which a finite polar Kerr effect develops in underdoped Y123 [Xia08]. These collective results were interpreted as strong evidence for broken time-reversal or inversion symmetry below $T^*$, possibly involving a novel current-loop order at $Q = 0$ [Varma97]. Subsequent muon spin rotation ($\mu$SR) studies, however, showed that the feature around $T^*$ may arise from a small amount (< 5%) of impurity phase [Sonier09], while the debate about the correct interpretation of the feature seen in the polarized neutron diffraction continues to rumble [Croft17, Bourges17]

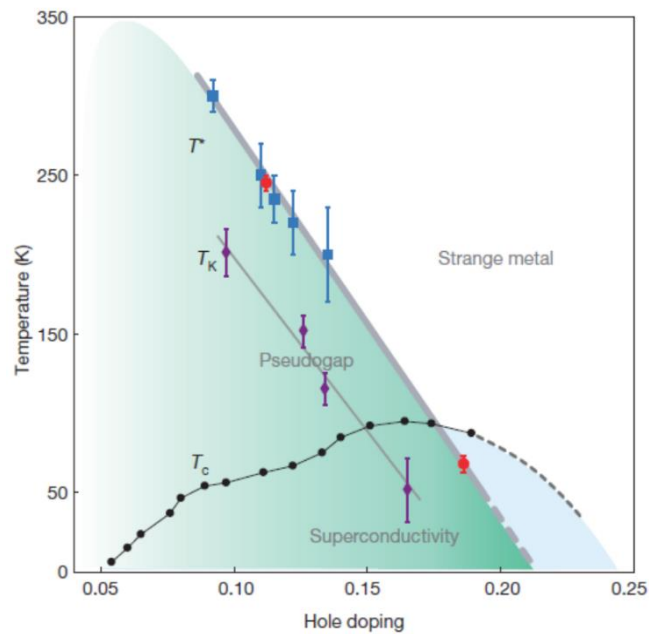

**Figure 6**: The phase diagram of $YBa_2Cu_3O_{6+\delta}$ showing the various ordering onset temperatures as determined by polarised neutron diffraction measurements (blue squares), resonant ultrasound (red circles) and Kerr rotation (purple diamonds) [Shekhter13].

Resonant ultrasound measurements on overdoped, fully oxygenated and underdoped Y123 showed evidence for a thermodynamic phase transition at $T^*(p)$ ascribed naturally to the opening of the pseudogap [Shekhter13]. Significantly, the measurements on fully loaded Y123 showed an onset temperature occurring below $T_c$ (see Figure 6), leading the authors to conclude that $T^*$ itself must collapse to zero at a doping level located inside the superconducting dome. Cooper *et al.* later analysed the data from Ref. [Shekhter13] and challenged its conclusions [Cooper14]. Firstly, the authors remarked that there is no other evidence for pseudogap formation in fully oxygenated YBCO ($p = 0.19$), though given the steepness of the $T^*$ line around the putative critical point, this may simply be a question of doping control (though later second harmonic generation measurements appeared to refute this [Zhao16]). More significantly, while the anomalies in resonant ultrasound across $T_c$ for both over- and underdoped samples are consistent with the specific heat, pressure derivatives of $T_c$, and the known values of the elastic constants, the observed changes in the mode frequencies at $T^*$ are incompatible with the lack of corresponding anomalies in the specific heat [Cooper14]. The authors thus concluded that the features in the resonant ultrasound data are due to thermally activated relaxation processes possibly involving oxygen ordering.

## 6. TRANSPORT AND THERMODYNAMIC PROPERTIES OF CUPRATES

The prevailing view of the pseudogap in hole-doped cuprates is one in which $T^*$ marks the onset of an ordered phase, involving some form of broken symmetry. By the same reasoning, the closure of the pseudogap at $p^*$ is a QCP where $T^*$ itself is suppressed to 0 K. A direct comparison with P-doped $BaFe_2As_2$ should in principle allow us to identify physical properties in a particular compound that are governed by its proximity to the QCP.

The pseudogap line has been tracked by a variety of techniques, including resistivity [Ando04], Raman spectroscopy [Sacuto13], neutron scattering [Fauque06], Kerr effect [Xia08] and others [Sato17, Shekhter13, Daou10, Zhao16]. On the one hand, $T^*$ is believed to represent a genuine feature in the phase diagram of cuprates. On the other, the variation in $T^*$ from the different experimental probes can be as large as 100 K for some doping values (see Fig. 6) [Ramshaw15], questioning the assumption of a single phase transition. Further problems are posed by the absence of an observable specific heat anomaly at $T^*$. The lack of a singularity in the specific heat is not a trivial point, since the opening of the pseudogap at $p^*$ has a profound influence on other physical properties (as discussed below). There may be phase transitions near the experimentally defined $T^*$ line, but significantly, they do not appear to have any effect on the entropy of the charge carriers because they are not seen in thermodynamic properties. It seems then that such phase transitions are not causing the pseudogap, but instead involve some electronic instability developing inside the pseudogap state.

Based on a series of detailed, differential heat capacity measurements, complemented by magnetic susceptibility and other bulk measurements, Tallon, Loram and co-workers have advocated a scenario in which the pseudogap reflects not a phase transition but an underlying energy scale $E_g$ which falls to zero beneath the SC dome. Their results and conclusions have not been accepted universally by the community however. This is in part due to the fact that the experiments themselves are extremely challenging and to date, the results have not been reproduced by any other group. Moreover, the electronic contribution to the heat capacity around room temperature constitutes less than 1 % of the total measured heat capacity and the subtraction of the dominant phonon term must rely on certain assumptions. Nevertheless, Loram and Tallon were able to deduce a number of important features relating to the pseudogap phenomenology from their analysis that have survived the test of time remarkably well, implicating both the power of the technique and the reliability of their analysis.

The first significant finding is the suppression of $E_g$ to zero at a doping level $p^* \sim 0.19$ inside the dome that is similar in all cuprate families, despite the marked differences in electronic structure (single, double and triple layers, van Hove singularities at different hole concentrations and different Fermi surface curvature). For a long time, it was argued from analysis of various spectroscopic measurements [Hufner08] that the pseudogap persisted to the edge of the dome. However, the community has now converged on the Tallon/Loram proposal, the alternative conclusion having been influenced, it seems, by the difficulty in distinguishing spectroscopically between an anti-nodal pseudogap and the pairing gap itself.

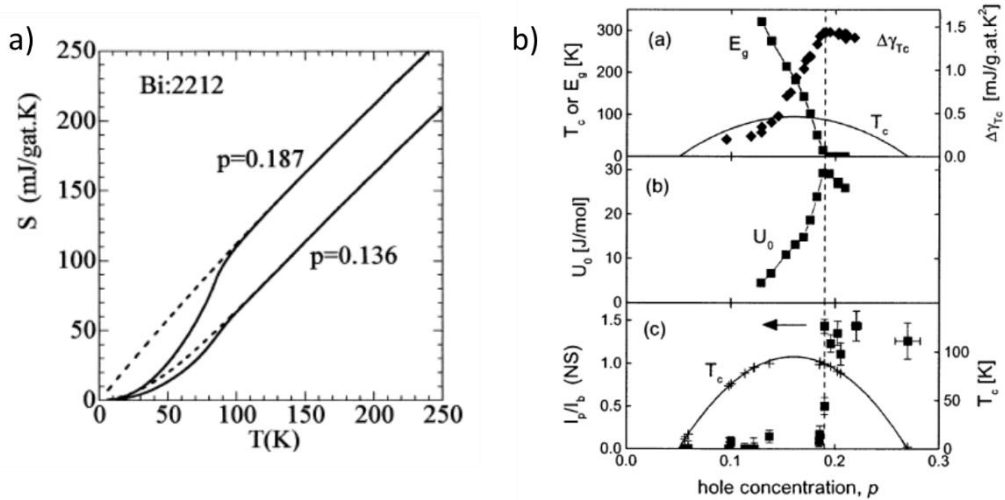

**Figure 7**: a) The electronic entropy of overdoped and underdoped Bi2212 [Loram01]. b) The doping dependence of (top panel) the pseudogap energy scale $E_g$ and the jump in the specific heat at $T = T_c$, (middle panel) the condensation energy $U_0$ and (bottom panel) the intensity of the quasiparticle peak near the Brillouin zone boundary extracted from ARPES studies of Bi2212 [Tallon03].

In 2011, He *et al.* [He11] presented a combined study of ARPES, Kerr effect, and time-resolved reflectivity in optimally doped Bi2201, all consistent with a mean-field-like vanishing of an order parameter at $T^*$. Inside the pseudogap regime, however, the as-measured electronic entropy does not indicate a conventional Fermi surface instability involving spin (charge) density wave or SC correlations. As shown in Fig. 7a, the entropies of over- and underdoped Bi2212 (obtained through an integration of the electronic specific heat) do not merge at $T^*$, but instead evolve in near parallel fashion with increasing temperature [Loram01]. This implies that the entropy lost at low temperatures (i.e. within the pseudogap phase) is not recovered at high $T$. This *non-states-conserving* nature of the pseudogap, at least within the energy range $\varepsilon_F \pm 100$ meV covered by Tallon and Loram's experiments, appears at odds with a mean-field-like vanishing of an order parameter at $T^*$.

The third key feature is illustrated in Fig. 7b. The collapse of the condensation energy $U_0$ below $p^*$ (middle panel) coincides with a sudden and marked drop in the specific heat jump at $T_c$, a measure of the pair density (top panel) [Tallon03]. A reduction in superfluid density below $p^*$ was first revealed by $\mu$SR [Uemura89]. Initially, the 'Uemura relation' – the proportionality between $T_c$ and the zero-temperature superfluid density $n_s(0)$ (or phase stiffness) – lent support to the phase fluctuation scenario in which phase coherence is destroyed but the amplitude of the order parameter remains finite. However, taking into account the specific heat data, one can infer from the Uemura relation that spectral weight that is first lost inside the pseudogap phase upon cooling is never recovered in the superconducting condensate.

The bottom panel of Fig. 7b shows that the ratio of the (coherent) quasiparticle peak to the incoherent background, as obtained by ARPES, also drops precipitously at $p^*$. This suggests that the pseudogap competes with superconductivity by depleting the low-energy electrons in the anti-nodal region (i.e. near the zone boundary) that would otherwise form pairs below $T_c$; in underdoped cuprates, only a reduced portion of the Fermi surface close to the nodal region contributes to the superfluid density. This is confirmed by polarized Raman studies that show electronic coherence persists down to low doping levels at the nodes (in the $B_{2g}$ channel), whereas anti-nodal quasiparticles (in the $B_{1g}$ channel) become incoherent below $p^*$ [LeTacon06]. Finally, the impact of the pseudogap on the superfluid density leads to a broad minimum in $\lambda_L^2(0)$ at $p^*$, in marked contrast to the sharp enhancement of $\lambda_L^2(0)$ in BaFe$_2$(As$_{1-x}$P$_x$)$_2$ that was interpreted earlier in terms of an enhanced effective mass near the QCP. Two messages must be taken from the above observations: i) the pseudogap is not a precursor superconducting state and ii) the nature of the quantum criticality in cuprates, if present, must be very different from that found in the iron pnictides.

We now turn to discuss the normal state transport properties, in particular the evolution of the in-plane resistivity $\rho_{ab}$ with temperature and doping. Studies above $T_c$ revealed the presence of a large V-shaped region in the phase diagram, flanked by $T^*$ on the low doping side and a 'coherence' temperature $T_{coh}$ (to be discussed in more detail below) on the other, terminating at $p^*$ where the two temperature lines merge [Naqib02, Ando04]. At first glance, this behaviour appears very similar to that shown in Figure 2a for $BaFe_2(As_{1-x}P_x)_2$ suggesting a similar scenario in which the non-Fermi-liquid resistivity arises due to the presence of a zero-temperature quantum phase transition, possibly associated with the reconstruction of the Fermi surface. However, while suppression of superconductivity by the application of a high magnetic field confirmed this conjecture in $BaFe_2(As_{1-x}P_x)_2$, it turns out that in cuprates, $\rho_{ab}(T)$ displays an altogether distinct behaviour.

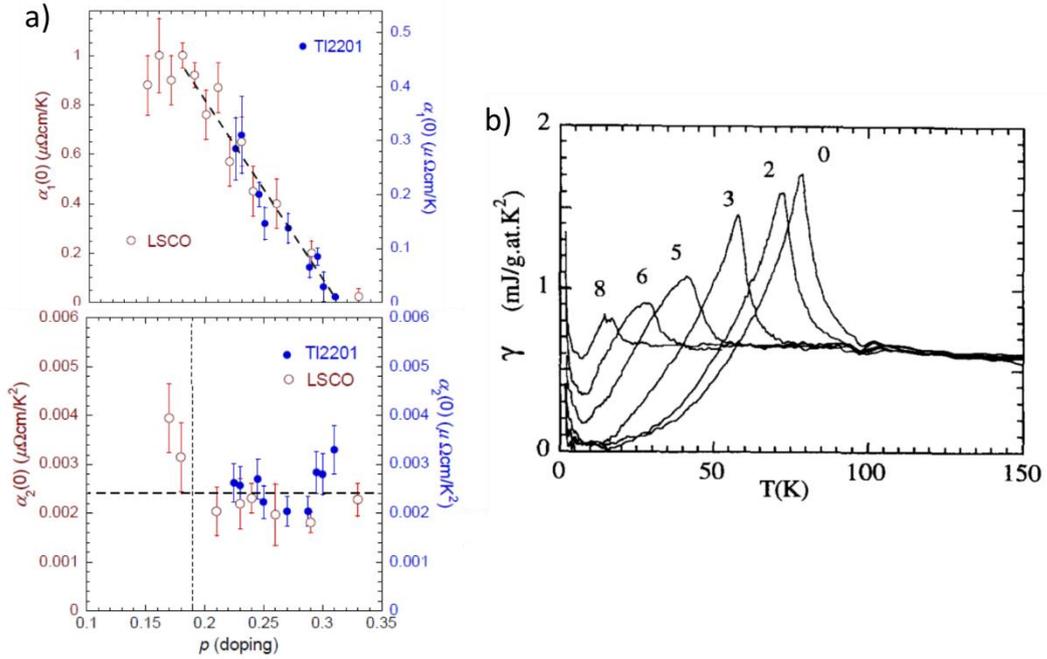

**Figure 8**: a) The evolution of the $T$-linear (top panel) and $T^2$-coefficient (bottom panel) as a function of hole doping are compared for Tl2201 (blue filled circles) and LSCO (red empty circles) [Hussey13]. b) Electronic specific heat coefficient of different Tl2201 overdoped samples. The number labels indicate samples with increasing doping levels [Loram94].

Cooper *et al*. used pulsed high magnetic fields to uncover the normal state behaviour of $\rho_{ab}(T)$ in LSCO down to 1.5 K across a broad range of dopings – from optimal doping $p$ = 0.17 to $p$ = 0.33 [Cooper09] (below optimal doping, the response is non-metallic [Boebinger96]). Over the entire doping range studied, $\rho_{ab}(T)$ could be decomposed into a combination of a $T$-linear and a $T^2$ component, with the $T$-linear term always dominating the resistivity at the lowest temperatures. This anomalous or 'extended' criticality contrasts markedly with the 'punctual' criticality observed in P-doped Ba122 [Analytis14] and in other candidate quantum critical systems such as YbRh$_2$Si$_2$ [Custers03] and indicates the presence of a quantum critical *phase* that persists over a finite doping range. This evolution of the resistive response in cuprates is depicted schematically in Figure 1b.

Fitting the resistivity as $\rho_{ab}(T) = \alpha_0 + \alpha_1 T + \alpha_2 T^2$, Cooper *et al.* found a monotonic scaling of $\alpha_1$ with $T_c$ on the overdoped side, reaching a peak or saturation value at $p^* = 0.19$. The $\alpha_2$ coefficient, meanwhile, showed no sign of the divergence expected near a QCP. As shown in the combined plot of Fig. 8a, the variation of $\alpha_1$ and $\alpha_2$ is the same in LSCO as in another single-layered cuprate Tl$_2$Ba$_2$CuO$_{6+\delta}$ (Tl2201) [Hussey13, Mackenzie96, Proust02]. Earlier angle-dependent magnetoresistance (ADMR) measurements in Tl2201 revealed that the two components in the resistivity may arise from two independent scattering mechanisms, one that is isotropic over the Fermi surface and quadratic in temperature and a *T*-linear term that is strongly anisotropic, peaking at the zone boundary and vanishing along the zone diagonal [Abdel-Jawad06, Abdel-Jawad07]. Interestingly this distinctly non-Fermi-liquid dependence of the electrical resistivity occurs in a region of the phase diagram where coherent fermionic quasiparticles persist, as revealed by the observation of quantum oscillations [Vignolle08], persist. The observation of fast oscillations in overdoped Tl2201 is particularly significant as it demonstrates that its electronic state is highly homogeneous; the effective carrier density varying by less than 0.1% over a distance equivalent to the cyclotron radius of order 200 nm [Bangura10]. The similar evolution of the resistivity in two distinct cuprate families with different Fermi surface topologies, electronic anisotropies, disorder levels and homogeneity provides strong evidence that this strange-metal behaviour is a generic property of all overdoped cuprates. Finally, the doping-independence of $\alpha_2$ (beyond $p^*$) is consistent with the observation, shown in Fig. 8b, that the normal state $\gamma$ (and $\Delta C/T_c$) in Tl2201 remains essentially constant upon approaching $p^*$ [Loram94].

To summarize, we present in the table below the contrasting behaviour of hole-doped cuprates and BaFe$_2$(As$_{1-x}$P$_x$)$_2$ in the vicinity of $p^*$. Clearly, their transport and thermodynamic responses reveal some fundamental differences between the two high-$T_c$ families, differences that have largely gone unnoticed. The absence of a divergence in the $T^2$ coefficient of $\rho_{ab}(T)$, or in $\gamma$ (just above the jump in the specific heat at $T_c$), combined with the persistence of the *T*-linear term in $\rho_{ab}(T)$ across the entire overdoped region, appears to be inconsistent with the assignment of the closing of the pseudogap to a conventional QCP. There is no question that broken symmetry states exist within the pseudogap phase. However, the lack of any specific heat anomaly at $T^*$ indicates that it is not appropriate to identify $T^*$ as the transition to the symmetry breaking phase whose associated fluctuations determine the dressing of quasiparticles beyond $p^*$. In order to reconcile these contradictions, a special sort of quantum criticality is required, one that is local in space, and so featureless in **k** [Keimer15], yet this idea conflicts with ARPES [Yusof02], Raman [Opel00] and ADMR [Abdel-Jawad06] that all show strong anisotropy in the self-energy and in the scattering rate. Moreover, the non-states-conserving nature of the pseudogap, as revealed by specific heat and magnetization measurements, is at odds with the conventional picture of an electronic instability near the Fermi level, where one expects the spectral weight to be recovered over an energy window of order the gap magnitude. Given these marked discrepancies with the conventional picture of quantum criticality, we consider below an alternative explanation for pseudogap formation in cuprates based on the concept of anisotropic quasiparticle decoherence.

|  | **Cuprates** | **Pnictides** |
|---|---|---|
| $1/\lambda^2$ at $p^*$ | Maximum | Minimum |
| $\Delta C/T_c$ near $p^*$ | Constant (on OD side) | Divergent (on both sides) |
| $\rho_{ab}(T)$ beyond $p^*$ | $\rho_{ab} \sim T\ (+\ T^2)$ | $\rho_{ab} \sim T^2$ |
| $A$ near $p^*$ | Constant (on OD side) | Divergent (on OD side) |

**Table 1**: Contrasting behaviour in the cuprates and pnictides at or near their putative quantum critical point $p^*$. The properties listed are: (i) the superfluid density, as determined by the magnetic penetration depth $\lambda$, (ii) the jump in the electronic specific heat $\Delta C$ at $T = T_c$, normalized by $T_c$, (iii) the $T$-dependence of the in-plane electrical resistivity $\rho_{ab}(T)$ at low $T$ and (iv) the magnitude of the $A$ coefficient of the $T^2$ resistivity.

# 7. EVOLUTION OF QUASIPARTICLE INTEGRITY IN CUPRATES: A GENESIS FOR THE PHASE DIAGRAM.

The preceding discussions lead us to conclude that the origin of the pseudogap is *not* related to an underlying quantum phase transition, at least not in the conventional sense as observed in Ba122. Certainly, the lack of dependency of *p\** on the details of the electronic structure in different cuprate families suggests that pseudogap formation is determined primarily by correlation-driven physics and by the degree of incommensurability away from half-filling, rather than by any particular feature of the Fermi surface topology. Mott transitions in general manifest themselves in two guises: the bandwidth-controlled transition at half-filling, tuned by the ratio of the on-site *U* and bandwidth *W* and the filling-controlled transition, tuned by electron doping $\delta$ away from half-filling. In the former, the approach to the Mott insulating state is characterized by a diverging enhancement in the quasiparticle effective mass [Brinkman70], and in the latter, by the formation of a spin gap and a vanishing density of states [Imada93].

The key physical assumptions required for the former behavior are the presence of strong correlations and the locality of correlation effects, reflected in a momentum independent self-energy $\Sigma = \mathrm{Re}\Sigma + \mathrm{Im}\Sigma$ (= $\Sigma' + \Sigma''$). In cuprates, as mentioned above, there is strong evidence for momentum dependence in $\Sigma(k)$. The anisotropy of the pseudogap, being strongest for the states near ($\pi$, 0) and absent for states along the zone diagonals, is mirrored in the anisotropy of the scattering rate in OD cuprates, as seen in the self-energy ($\Sigma''$) determined by ARPES [Yusof02], and in the quasiparticle scattering rate as deduced, for example, through Raman spectroscopy [Opel00] or ADMR [Abdel-Jawad06]. In many respects, the scattering rate anisotropy that develops in the overdoped regime can be seen as a precursor of the pseudogap regime and thus, it is worthwhile to explore in more detail the evolution of the electronic state from the correlated Fermi-liquid state beyond the SC dome [Nakamae03] into the strange metallic and pseudogapped states at intermediate dopings.

Kaminski and co-workers carried out an ARPES study of the anti-nodal spectral function in overdoped cuprates as a function of temperature and doping and found that the quasiparticle integrity, as deduced from the presence of a peak in the coherent part of the single-particle spectral function, disappears with increasing *T* [Kaminski03] Moreover, this loss of quasiparticle integrity for states at the zone boundary was found to coincide with the onset of *T*-linear resistivity. The key experimental results are reproduced in Figure 9.

This striking correspondence suggests that the strange metal state above $T_{coh}$ in Figure 9d is populated by incoherent excitations or non-quasiparticle states. What this particular study does not confirm however is whether states in other regions of the Fermi surface also suffer the same fate, i.e. whether the coherent / incoherent crossover is $k$-selective or occurs simultaneously over the entire Fermi surface. Indeed, the current state-of-the-art paints a conflicting picture regarding the extent of the phase diagram over which quasiparticle integrity, e.g. at the nodes, survives. Nevertheless, a number of ARPES studies have shown that the low-energy nodal response is indeed sharp enough to support quasiparticle excitations (with an inverse lifetime that scales quadratically in energy [Kordyuk05, Koralek06, Reber15, Fatuzzo14]) at least down to optimal doping, implying that the nodal quasiparticles preserve their integrity over a much wider region of the phase diagram. This dichotomy has also been seen in the collective response, e.g. via Raman studies [Opel00, LeTacon06]. It is surprising that such a qualitative change in an angle-integrated quantity - namely the crossover to a $T$-linear in-plane resistivity at $T_{coh}$ - coincides with the loss of quasiparticle integrity at the anti-nodes.

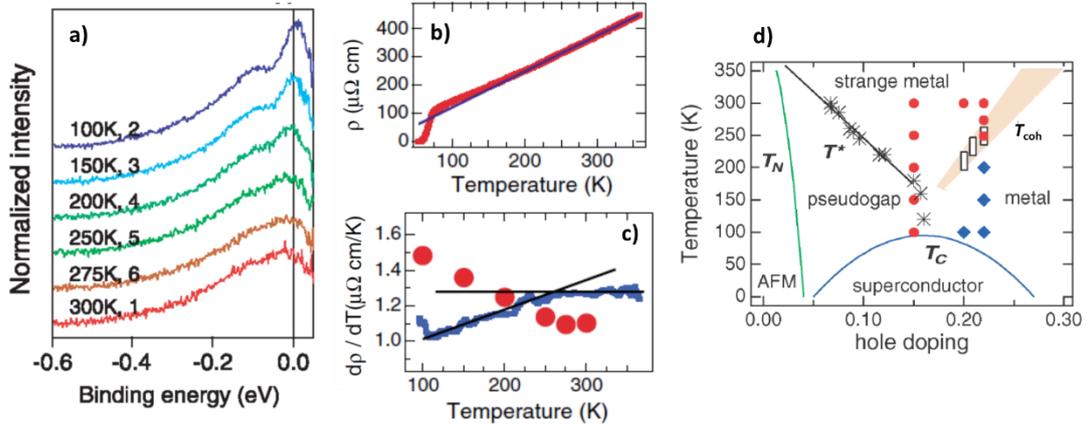

**Figure 9:** a) Single-particle spectral function at $(\pi, 0)$, divided by the Fermi function, for OD Bi2212 ($p = 0.23$) at various temperatures. b) $\rho_{ab}(T)$ curve for the same sample. The blue line denotes a linear fit to the high temperature data. c) d$\rho_{ab}$/d$T$, with black lines guides to the eye. Red dots are the ARPES intensities obtained from (a). d) Resultant phase diagram. The 'strange metal' region is defined by the departure from $T$-linear resistivity. Red and blue symbols correspond to the absence or presence of coherent ARPES peaks respectively [Kaminski03].

Figure 10a shows the phase diagram for LSCO extracted from d$\rho_{ab}$/d$T$ for a range of dopings [Hussey11] with $T_{coh}$ and $T^*$ similarly defined as the temperatures below which $\rho_{ab}(T)$ deviates from its high-$T$ $T$-linear behaviour. Within the experimental uncertainty, $T_{coh}$ appears to collapse to zero at or around the same doping level at which the pseudogap opens. Figure 10b compares the doping dependence of $\alpha_1(\infty)$, the coefficient of the $T$-linear term in the high-$T$ strange metal phase of LSCO (filled diamonds) and $\alpha_1(0)$, the coefficient of the corresponding low-$T$ $T$-linear term reproduced from Figure 8a (filled circles). The doping dependence is clearly different. Above $p^*$, $\alpha_1(\infty)$ attains a constant value of ~ 1.0 $\mu\Omega$cmK$^{-1}$, suggesting some sort of fundamental limit. For a two-dimensional Drude metal, $\alpha_1(\infty)$ can be expressed as:

$$\alpha_1(\infty) = \frac{d\rho_{ab}}{dT} = \frac{2\pi\hbar d}{e^2 k_F v_F} \frac{d(1/\tau)}{dT}$$

where $d$ is the interlayer spacing, $v_F$ is the Fermi velocity and $k_F$ the Fermi wave vector. For LSCO, this corresponds to a momentum-averaged transport scattering rate $\hbar/\tau \sim \pi k_B T$. Moreover, given that the anisotropic scattering rate varies approximately as $\cos^2 2\phi$ within the plane ($\phi$ being the angle between the $k$-vector and the Cu–O–Cu bond direction) [Abdel-Jawad06], this is equivalent to $\hbar/\tau \sim 2\pi k_B T$ for states near ($\pi$, 0). This is the same threshold value that was introduced in Section 3 (the Planckian dissipation limit) and defines an upper bound beyond which quasiparticle coherence is claimed to be lost and charge transport becomes driven by diffusion of charge and energy rather than by momentum relaxation. A similar bound can be deduced for the *single*-particle scattering rate $\Gamma$ from the spectral linewidth analysis of ARPES measurements in, for example, Nd-LSCO [Matt15].

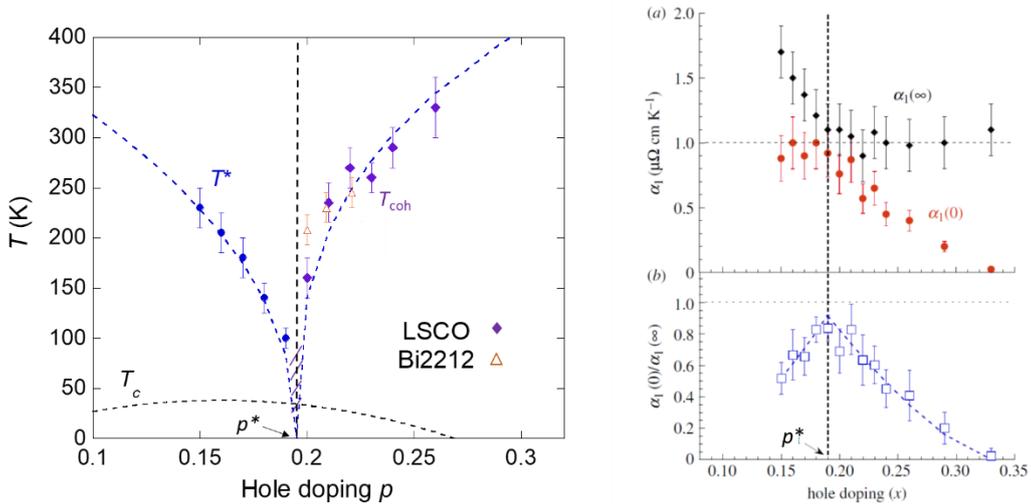

**Figure 10:** a) Temperature versus doping phase diagram of La$_{2-x}$Sr$_x$CuO$_4$ (LSCO) as extracted from d$\rho_{ab}$/d$T$. The labels $T^*$ and $T_{coh}$ define the region of $T$-linear resistivity above and below $p_{crit}$ = 0.19. The open diamonds are the $T_{coh}$ values for Bi2212 from Ref. [Kaminski03]. b) Top panel: Doping dependence of $\alpha_1(\infty)$ (filled diamonds) and $\alpha_1(0)$ (filled circles) for LSCO. Bottom panel: Ratio of $\alpha_1(0)/\alpha_1(\infty)$ as a function of Sr content [Hussey11].

Given the evolution of the *T*-linear resistivity within the strange metal phase, it would appear that the Planckian limit is playing a significant and hitherto unrecognized role, not only in the development of the pseudogap, but also in shaping much of the cuprate phase diagram. Above $T_{coh}$, the temperature derivative of the quasiparticle scattering rate within the strange metal phase is at or close to the Planckian limit. Fig. 10a shows that $T_{coh}$ itself collapses towards zero at the critical doping level *p**. At the same time, $\alpha_1(0)$, the coefficient of the low-*T* *T*-linear resistivity approaches the same limit. Thus at *p* = *p**, the scattering rate (for states near ($\pi$, 0)) is close to the Planckian limit even at zero temperature, and thus below *p**, one must surmise that the anti-nodal quasiparticles are never coherent. In this regard, we note that scanning tunnelling spectroscopy measurements also find that below *p**, quasiparticle interference – a definitive signature of quasiparticle coherence – vanishes for those states near the zone boundaries even at the lowest temperatures [Kohsaka08].

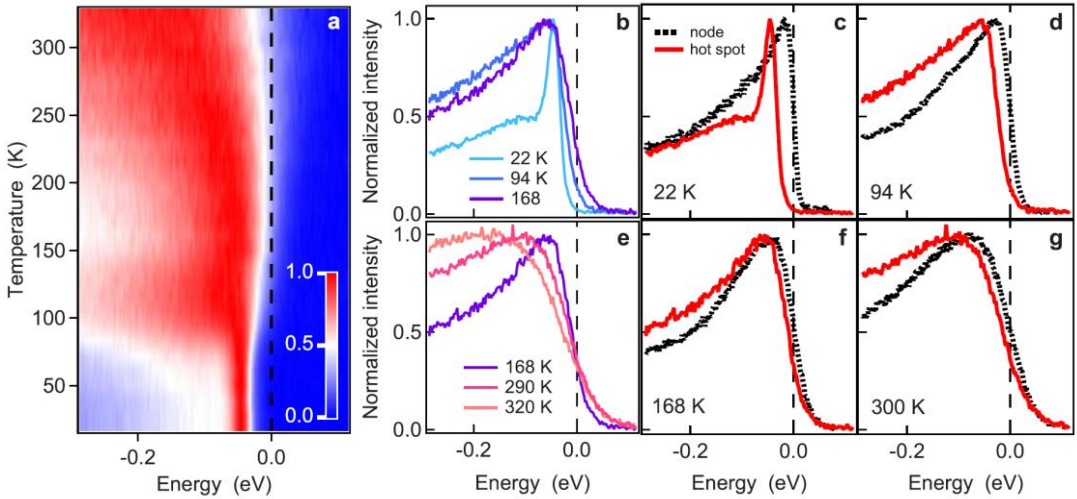

**Figure 11:** a) Temperature dependence of momentum-integrated EDCs measured at a hot spot. The pseudogap is seen as a shift of the LEM (white color close to the Fermi level). Separate EDCs are shown in panels (b–g): as compared to each other (panels b and e) and to the similar EDCs measured each for the same temperature but along the nodal direction (panels c, d, f, g) [Kordyuk09].

The pseudogap has often been defined in terms of an energy scale $\Delta_0$, an energy window over which states are removed from the Fermi level $E_F$. In this case, however, one might expect that at temperatures larger than $\Delta_0$, the states will be restored, especially when $\Delta_0$ is low. In reality, the pseudogap never actually closes – otherwise the entropy lost below *T** would be recovered. This has been shown explicitly by Kordyuk *et al.* [Kordyuk09] who performed ARPES on an underdoped Bi2212 sample with high energy resolution. The key spectra, shown in Fig. 11, were measured along the cut of the Brillouin zone (BZ) through the antiferromagnetic hot spots (where the pseudogap should be largest) and compared with similar spectra obtained around the nodes. The energy distribution curves (EDCs) were integrated over a finite momentum range around $k_F$ to ensure that the leading edge midpoint (LEM) of a *non-gapped* spectrum always remained at the Fermi level. Kordyuk *et al.* found that the pseudogap persists over the whole temperature range up to 300 K, i.e. above the temperature scale at which the pseudogap itself is expected to close. Intriguingly, the gap value was also found to be non-monotonic, reaching a finite minimum value at about 170 K and increasing on both sides. The shift of the LEM also appeared to be correlated with the width of the spectra: the closer the LEM is to the Fermi level, the narrower the spectrum.

While other ARPES data have been interpreted in terms of a closing of a pseudogap with increasing *T* [Hashimoto10, He11, Kondo13], it is important to reiterate a couple of salient points. Firstly, in the absence of a spectral gap, Re$\Sigma$ = 0 at $k = k_F$, and the spectral intensity is simply a Lorentzian of width Im$\Sigma$, when approximating Im$\Sigma$ by a constant $\Gamma$, which is effectively a measure of the quasiparticle scattering. As scattering increases, the linewidth broadens and the peak amplitude is lowered. In this way, a metal can lose its coherence. It is difficult to determine a true gap when $\Delta \ll \mathrm{Im}\Sigma(\Delta)$. Secondly, pseudogap closure is at odds with bulk specific heat, NMR and magnetization data [Tallon03]. If the spin susceptibility $\chi$, for example, is constant above $T_c$, the temperature dependence of the product $\chi T$ will be linear through the origin. In the presence of a *d*-wave-like pseudogap with gap energy $E_g$, the product $\chi T$ is depressed downwards due to the reduced DOS. If $E_g$ is temperature independent $\chi T$ is displaced downwards in parallel fashion. If the pseudogap opens at *T\**, however, then $\chi T$ returns to the parent linear curve, meeting it at *T\**. The latter situation has not been observed for any cuprate studied thus far. Instead, with progressive underdoping, these curves form a series of downward-moving parallel lines representing the increasing pseudogap energy and mimicking the behaviour seen in the entropy *S* in Y123 in Fig. 7a [7].

The depletion of low-energy spectral weight with increasing temperature is reminiscent of what is observed in the optical reflectivity of cuprates and other 'bad metals'. Indeed, the cuprates were the first material to be recognized as bad metals, their electrical resistivity increasing with temperature seemingly without bound, reaching values that were incompatible with the coherent propagation of Bloch waves [Gurvitch87]. It appears then that both bounds (MIR and Planckian) are violated in cuprates, the latter being correlated with the loss of quasiparticle coherence at the Brillouin zone boundary and the opening of the pseudogap. The key ingredient that still needs to be incorporated into the overall picture, however, is the *k*-space dichotomy that creates a co-existence of coherent and incoherent states at different loci on the (underlying) Fermi surface whose own ratio evolves with temperature. With this in mind, we introduce here (Figure 12) a new proposal for what may ultimately define and shape the phase diagram of hole-doped cuprates.

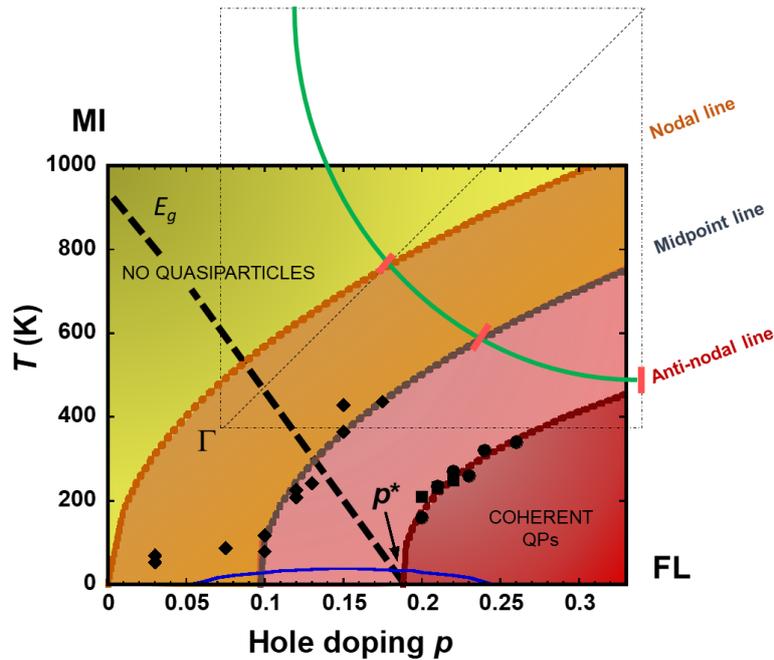

**Figure 12:** Loss of quasiparticle integrity across the cuprate phase diagram. The thick dashed lines define three representative lines, each one specifying the temperature beyond which quasiparticle integrity is lost at a particular locus on the underlying Fermi surface; at the zone boundary (anti-nodal line), along the zone diagonal (nodal line) and at a point halfway between the two (midpoint line). The solid squares and circles represent $T_{coh}$ for Bi2212 [Kaminski03] and LSCO [Hussey11] respectively. The diamonds represent $T_0$ determined by AIPES on Bi2212 (Hashimoto09) (see text for details).

The figure shows the temperature vs. doping phase diagram superimposed on the upper-right quadrant of the first Brillouin zone, the solid green line representing schematically a 2D projection of the underlying Fermi surface. The dotted curves define the threshold for the loss of quasiparticle integrity at three key loci on the Fermi surface; at the Brillouin zone boundary (anti-nodal line), along the zone diagonal (nodal line) and at a point halfway between the two (midpoint line). In reality, there will be a cascade of such lines, presumably parallel to each other, their boundary defined by the two extremes. Our main quantitative claim is that each line defines the point in the temperature vs. doping landscape at which the quasiparticle scattering rate reaches the diffusive bound (defined here as $\hbar/\tau = 2\pi k_B T$) for that particular location within the Fermi quadrant.

Hashimoto *et al.*, performed systematic temperature and doping dependent angle-*integrated* photoemission (AIPES) measurements on LSCO and derived a crossover temperature or equivalent coherence temperature $T_0$ from a shift in the peak positions in the derivatives of the AIPES data [Hashimoto09]. In the underdoped region, $T_0$ was found to be smaller than $T^*$, implying that with increasing temperature, the quasiparticles on the Fermi arc start to lose coherence well below $T^*$, and in the temperature range $T_0 \leq T \leq T^*$, the pseudogap opens on the 'incoherent' portion of the Fermi surface. According to the authors, the apparently metallic transport at higher temperatures (> $T_0$) in the underdoped region is governed by incoherent hole carriers on the Fermi arc/surface. In this region of the phase diagram, it is likely that both the MIR and diffusive bounds are violated, removing coherent quasiparticle weight in the process. The estimates for $T_0$ obtained from the AIPES measurements are superimposed on the phase diagram in Fig. 12. As we can see, the derived values coincide well with the coherence line for the midpoint between the node and the antinode, consistent with it being the average of the response from the entire Fermi surface. Also plotted in Fig. 12 are values for $T_{coh}$ for the antinodes estimated from the ARPES measurements of Kaminski *et al.* [Kaminski03] and the corresponding boundary of the strange metal phase deduced from in-plane resistivity [Hussey11].

In order to test this hypothesis of a cascade of coherent-to-incoherent boundaries, one would need to carry out a systematic $T$-, $p$- and $k$-dependent ARPES study on cuprates beyond $p^*$. Such a study would, we believe, be relatively straightforward though unavoidably time-consuming. A more quantitative determination of the precise bound responsible for any loss of quasiparticle integrity, however, may be more challenging but nonetheless achievable.

## 6. DISCUSSION AND CONCLUSIONS

In our search for guiding principles and universality, it is hard not to be intrigued or seduced by the similarities between different families of superconductors. The similarities in the cuprates and pnictides are indeed numerous and taking a cursory glance at their respective phase diagrams, one might be led to the conclusion that $p^*$ defines a quantum critical end point in the former and that the pseudogap itself is associated with some form of ordering that sets in at $T^*$. Many candidate order parameters have been proposed, including nematic, antiferromagnetic, charge and bond-centred *d*-density wave order. One of the central claims of this article is that all of these broken symmetries are merely instabilities of the pseudogap phase and play no role in the emergence of the pseudogap *per se*. Instead, we have introduced an empirical picture in which strong electron correlation effects within the $CuO_4$ plaquette emerge in the strongly overdoped region of the phase diagram and first induce an anomalous (linear in temperature and frequency) and anisotropic (Umklapp) scattering on initially coherent quasiparticle states that grows in intensity with decreasing doping and finally crosses a coherence bound at $T_{coh}$. This interaction then begins to consume the quasiparticle weight in a manner that spreads progressively across the underlying Fermi surface with further decrease in the effective carrier number, leading to the formation of Fermi arcs whose own extent gradually disappears as the Mott state is approached [Liu17].

High-$T_c$ superconductivity in the cuprates is one of the most spectacular examples of a doped Mott insulator in which superconductivity emerges from an unconventional normal state. The scenario reported here is essentially in the spirit of theoretical models based on a doped Mott insulator [Lee06] that predict the existence of the coherent spectral weight in the single-particle excitation spectrum that is proportional to *x*. How (and whether) coherent quasiparticles form in a lightly doped Mott insulator has been a key question in the physics of strongly correlated electron systems. The reduced mobility of the charge carriers as the Mott insulating state is approached will certainly provide the right environment for instabilities to flourish, particularly at low-temperatures where thermal fluctuations are frozen out, but it is important to acknowledge that none of these instabilities can be the driving force for pseudogap formation [Peli17]. The key additional ingredients inferred from the analysis presented here are the variation of the coherent / incoherent boundary around the underlying Fermi surface and the role of the Planckian dissipation limit in defining the location of *p\**.

It should not be forgotten that there are a number of 'conventional' QCPs buried within the cuprate phase diagram; the vanishing of long-range antiferromagnetism at *p* ~ 0.05 (in hole-doped cuprates) and at *p* ~ 0.15 (in electron-doped cuprates), the end-points of the intermediate-range charge ordered state (in high magnetic fields) at *p* ~ 0.09 and 0.16 (as evidenced by the divergent effective masses deduced from quantum oscillation experiments [Ramshaw15]) and possibly too, at the end of the SC dome. It is also true that critical fluctuations associated with such order can drive the quasiparticle states incoherent. Importantly, however, none of these 'conventional' quantum critical points coincide with $p^*$.

It should be emphasized too that the role of magnetism has not been explicitly considered here. The undoped phase is of course a Heisenberg antiferromagnet with a large exchange coupling $J_{AF}$ (on the scale of 100 meV) and it is widely assumed that the pseudogap represents some sort of crossover region in which AFM correlations with moderate correlation length lead to the formation of singlets between sites. While this picture may be appropriate for describing the cuprates at very low doping, it is not clear how such a scenario can survive out to 20 % doping nor give rise to such a sharp cut off of the pseudogap at $p^*$, given that the spin fluctuation spectrum evolves so smoothly over the entire phase diagram. And to reiterate, without strong evidence for a genuine QCP at $p^*$, it is the opinion of the authors that the only viable reason why $p^*$ is (a) so well-defined and (b) has such a similar value for all the various hole-doped cuprates, is that it reflects the presence of a universal bound.

Returning to our initial motivation – the search for guiding principles that could optimize superconductivity – there are three final points to make. Firstly, near a QCP, quantum critical fluctuations become long range, but at the same time, their frequency scale diminishes. This is not good for superconductivity. Nevertheless, critical fluctuations may enhance the bosonic coupling strength and so produce stronger Cooper pairing as the fluctuations themselves soften. In cuprates, superconductivity is optimized at $p = p^*$, though the pairing strength, as reflected in the magnitude of the superconducting gap is not (it peaks at optimal doping, $p$ = 0 .16). This appears to reaffirm the notion that the pseudogap end point is not linked to a vanishing (bosonic) energy scale. Secondly, quantum fluctuations can also enhance superconductivity by increasing the normal state energy and thus making superconductivity more energetically favorable. However, this so-called kinetic-energy-driven superconductivity, if it does exist in the cuprates, appears only below $p^*$ and is certainly not maximized at this doping level.

Finally, within the scenario presented in this review, $T_c$ rises from zero at the edge of the SC dome (on the overdoped side) as the anomalous ($T$, $\omega$ linear) anisotropic interaction emerges and grows in magnitude. The elevated $T_c$ values in cuprates imply that the pairing interaction itself is particularly strong. However, at $p = p^*$, the interaction becomes so strong that it starts to destroy the very quasiparticle states required to form the pairing condensate. This 'Catch-22' scenario proffers one possible route to optimize superconductivity in a suitably modified material. As the doping is reduced further towards the Mott state at half-filling, one expects this anisotropic interaction to continue to intensify. To enhance superconductivity further, one might consider a scenario in which the strength of the interaction is reserved, while at the same time, its magnitude is prevented from exceeding the upper bound that causes the quasiparticle states to decohere. One way to achieve this is to create the right structural environment in which the anisotropy of the same interaction is reduced or smeared out. Which specific modifications of the $CuO_4$ plaquette could allow this enhancement in $T_c$ to be realized remains to be found, but the journey itself could potentially be very rewarding.

The key issues that remain to be resolved are the origin of the *T*-linear resistivity within the strange metal regime, its associated anisotropic scattering rate and the interaction that gives rise to it. Within a Boltzmann-type approach, the *T*-linear resistivity is a low-energy effect (with a form set by the density of states of the associated boson) though its coefficient may be determined by high-energy physics. The extended range of the *T*-linear resistivity in overdoped cuprates, however, would require a very specific form of boson, one with an energy scale that does not collapse at a singular point in parameter space. Without evidence for such a boson, it may be necessary to look beyond our conventional approaches to many-body physics and consider models of charge dynamics that are driven purely by high-energy processes.